\documentclass{ws-p9-75x6-50}

\begin{document}

\title{A Cosmological Window on Trans-Planckian Physics}

\author{J\'er\^ome Martin}

\address{Institut d'Astrophysique de Paris CNRS,\\
98 boulevard Arago 75014 Paris, France\\
E-mail: jmartin@iap.fr}  

\author{Robert H. Brandenberger}

\address{Department of Physics, Brown University,\\
Providence, RI 02912, USA\\
E-mail: rhb@het.brown.edu}


\maketitle

\abstracts{In the framework of inflation, scales which nowadays 
correspond to large scale structures were smaller than the Planck 
length at the beginning of inflation. Therefore, measurements 
of CMBR anisotropy or surveys of galaxies and of clusters of galaxies 
could help us to probe physics beyond the Planck scale. In these 
proceedings, we study how the inflationary observables depend on 
trans-Planckian physics. It is found that there exist cases where 
adiabaticity can be violated and the primordial spectrum of the 
quantum cosmological perturbations modified. These cases 
correspond to rather drastic alterations of the dispersion 
relation. There exist also modifications of the standard physics 
which leave the power spectrum unchanged and therefore 
which are compatible with the data. In this sense, 
the presently available observations already constrain 
trans-Planckian physics.}

Since the recent observations of the Cosmic Microwave Background 
Radiation (CMBR) anisotropy and the detection of the first acoustic 
peak at an angular scale of $\simeq 1^{0}$ by the BOOMERanG and 
MAXIMA-1 experiments \cite{B98a,MAX1a}, it has become likely that 
inflation is the correct theory of the very early Universe. If so, this 
opens the interesting possibility to test some aspects of the 
physics beyond the Planck scale. Indeed, in most of the 
realizations of the idea of inflation, the 
phase of accelerated expansion lasts so long that present 
cosmological scales were below the Planck length at the beginning 
of inflation. Therefore, any alterations of the standard physics at 
extremely small scales could have direct consequences on what we observe 
today. Among the observables which would be changed is the power spectrum 
of the primordial quantum fluctuations. The goal of these proceedings is 
to study under which conditions the power spectrum can be modified 
by a change of the physics below the Planck length.
\par
As first discussed in Refs. \cite{Unruh,Brout,HB,CJ}, a similar problem 
exists in black hole physics. However, in this context, there exist deep 
thermodynamical reasons which render the derivation of the 
Hawking radiation robust to modifications of the trans-Planckian 
physics. For inflation, these reasons no longer exist (we 
recall that the power spectrum is no longer a thermal one) and therefore 
it would be incorrect to believe that this question is just a trivial 
technical extension of what has already been done in the context of 
black hole physics.
\par
Since a theory of quantum gravity is not available nowadays, the question 
arises as to which kind of modifications should be implemented in the 
theory. It was suggested by Unruh \cite{Unruh} that a generic way to do it is to 
consider non standard dispersion relations of the form $\omega =F(k)$ which 
reduce to the standard one for large scales, i.e. $F(k)\simeq k$ when $k$ 
is smaller than a cutoff $k_{\rm C}$ corresponding to the 
characteristic scale $l_{\rm C}\simeq 1/k_{\rm C}$. $l_{\rm C}$ could be, for 
example, the Planck scale. A theory of quantum gravity should provide us 
with the shape of $F(k)$ for large values of $k$. In absence of such a 
theory, we will consider different {\it ad hoc} shapes for $F(k)$. In 
cosmology, the physical wave number 
$k$ is related to the comoving one $n$ by $k=n/a(\eta )$ where $a(\eta )$ 
is the scale factor. We see that, in the context of cosmology, we are led 
to introduce time-dependent dispersion relations, namely 
$\omega _{\rm phys}=F[n/a(\eta )]$.
\par
The choice of the initial state also needs to be discussed 
carefully. Since the dispersion relation is modified, when 
the wavelength of a mode is smaller than the characteristic 
length, the mode does not behave as a free wave and the usual 
vacuum state cannot be chosen. However, in 
Refs. \cite{MB,BM}, it was shown that a direct generalization 
of the vacuum can be found and consists 
in the state which minimizes the energy density initially. In 
these proceedings, all the results are established under 
the assumption that the initial state is this ``minimizing 
energy state''\footnote{In Refs. \cite{MB,BM} another initial 
state was considered. Of course, only the ``minimizing energy state'' 
is physical and the purpose of introducing this other state was just 
to test the dependence of the final result on the initial 
conditions.}.
\par
In order to know whether a change in the spectrum is expected, one 
has to compare the characteristic time scale of evolution of the 
physical frequency, 
$(1/\omega _{\rm phys})({\rm d}\omega _{\rm phys}/{\rm d}\eta )$ 
with the Hubble time ${\cal H}\equiv a'/a$ in the regime where the 
physical wavelength is smaller than the characteristic scale, 
$\lambda /l_{\rm C}\ll 1$. For this purpose, it is convenient to 
define an ``adiabaticity coefficient'' by
\begin{equation}
\label{defalpha}
\alpha (n,\eta )\equiv \Biggl \vert \frac{{\cal H}}{\frac{1}{\omega _{\rm phys}}
\frac{{\rm d}\omega _{\rm phys}}{{\rm d}\eta }}\Biggr \vert .
\end{equation}
When $\alpha \gg 1$, the effective physical frequency can be considered 
as constant and the solution of the modified mode equation has time 
to adjust itself to track the standard solution. In this case, we have 
adiabaticity and no change is expected. On the contrary, if $\alpha \simeq 0$ 
adiabaticity is violated and some alterations of the standard power 
spectrum is likely to appear. In the standard case, we have $\alpha =1$ since 
the time scales of evolution of the physical frequency and of the 
Hubble time are the same.
\par
In Refs. \cite{MB,BM}, we have considered the following dispersion 
relations 
\begin{equation}
\label{disp}
\omega =k_{\rm C}\tanh ^{1/p}\biggl[\biggl(\frac{k}{k_{\rm C}}\biggr)^p\biggr], 
\quad \omega =k\sqrt{1+b_m\biggl(\frac{k}{k_{\rm C}}\biggr)^{2m}},
\end{equation}
which were first introduced in Refs. \cite{Unruh,CJ}. In the second of 
these equations, 
$m$ and $b_m$ are arbitrary numbers and $b_m$ can be either positive or 
negative. These dispersion relations are displayed in Fig. (\ref{disp}).

\begin{figure}[t]
\epsfxsize=25pc 
\epsfbox{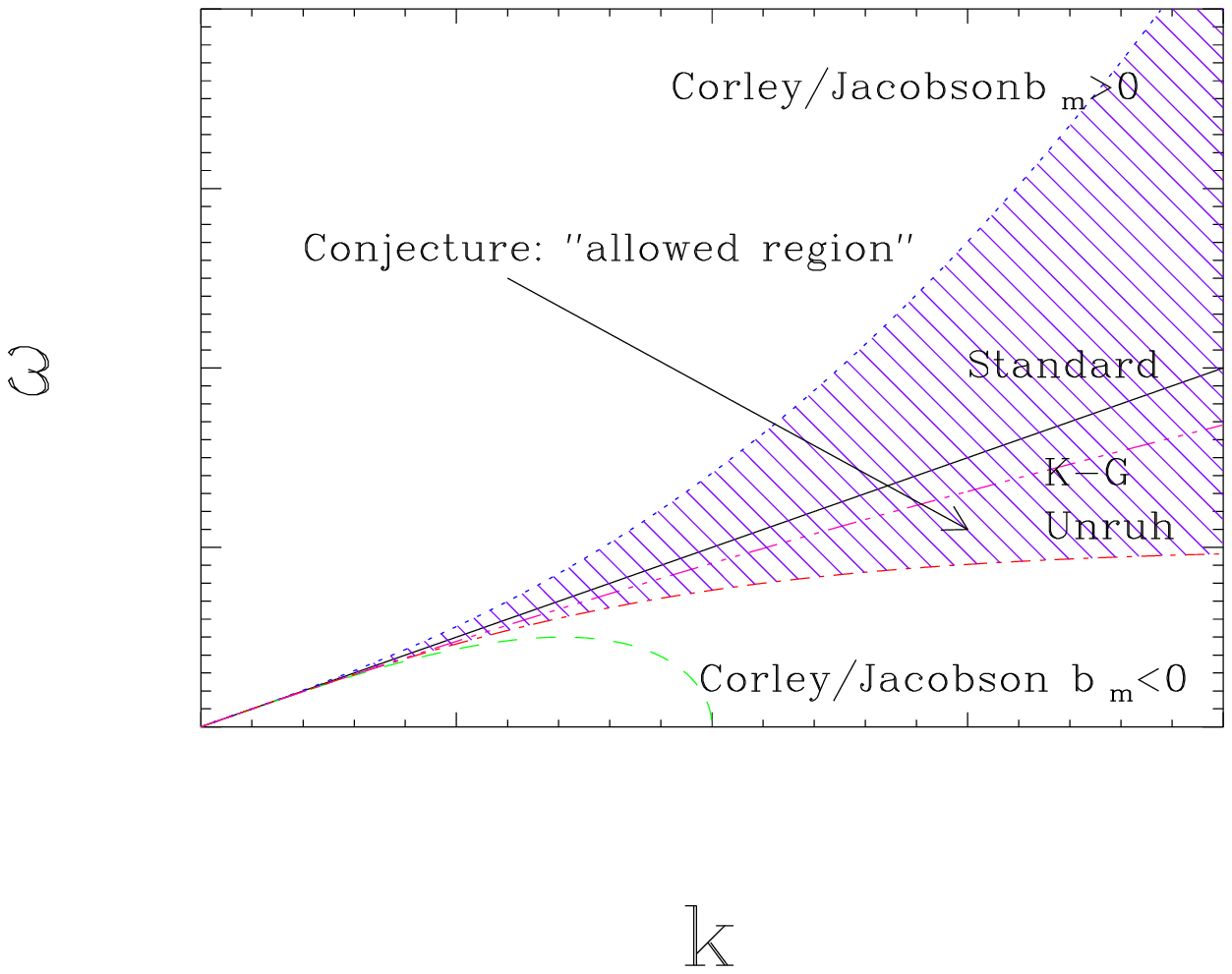} 
\caption{Different dispersion relations \label{disp}}
\end{figure}

\begin{figure}[t]
\epsfxsize=25pc 
\epsfbox{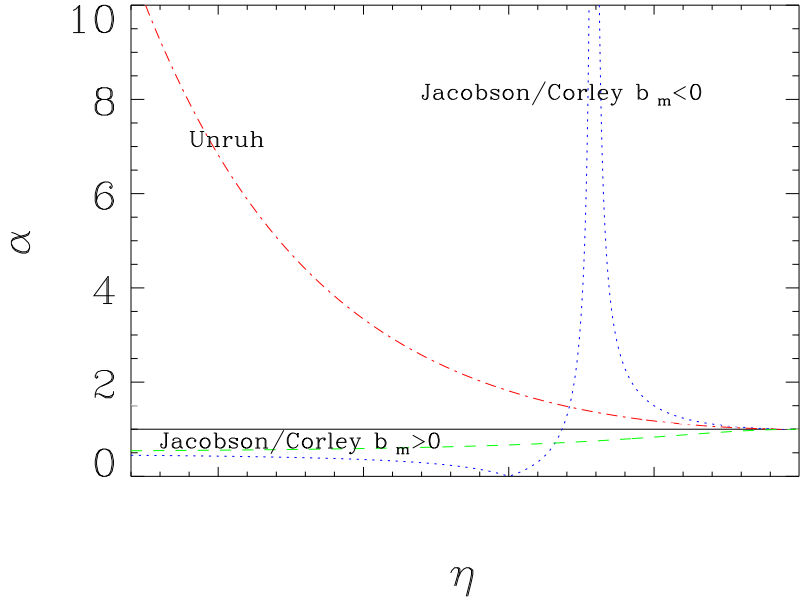} 
\caption{Different adiabaticity coefficients \label{al}}
\end{figure}

The adiabaticity coefficient can be computed exactly once a 
model of inflation is chosen. For simplicity, we consider power-law 
inflation where the scale factor is given by $a(\eta )=l_0|\eta |^{1+\beta }$, 
where $\beta \le -2$. $\beta =-2$ corresponds to the De Sitter 
spacetime. For the Unruh dispersion relation, one finds
\begin{equation}
\label{alu}
\alpha _{\rm U}(n,\eta )=\biggl(\frac{l_{\rm C}}{\lambda }\biggr)^{-p}
\sinh \biggl[\biggl(\frac{l_{\rm C}}{\lambda }\biggr)^p\biggr]
\cosh \biggl[\biggl(\frac{l_{\rm C}}{\lambda }\biggr)^p\biggr].
\end{equation}
When $l_{\rm C}/\lambda \ll 1$, one has $\alpha _{\rm U}=1$. On 
the contrary, if $l_{\rm C}/\lambda \gg 1$ then 
$\alpha _{\rm U}\rightarrow \infty $ 
and adiabaticity is satisfied. Therefore, we expect no change for 
the Unruh dispersion relation. This is confirmed by the results 
of Refs. \cite{MB,BM,Nie}. For the Corley-Jacobson dispersion 
relation the result reads
\begin{equation}
\label{acj}
\alpha _{\rm CJ}(n,\eta )=\frac{1+b_m(l_{\rm C}/\lambda )^{2m}}
{1+(b_m+1)(l_{\rm C}/\lambda )^{2m}}.
\end{equation}
One checks again that in the case where $l_{\rm C}/\lambda \ll 1$, 
one has $\alpha _{\rm CJ}=1$. If $l_{\rm C}/\lambda \gg 1$, the 
adiabaticity coefficient tends to the constant $1/(m+1)<1$. If 
$b_m<0$, $\alpha _{\rm CJ}$ goes to zero at the point where 
$\omega =0$ and adiabaticity is violated. In addition, it blows up at the 
point where ${\rm d}\omega /{\rm d}\eta =0$. The evolution of the 
adiabaticity coefficients is displayed in Fig. (\ref{al}).

The case in which a difference is obtained can be viewed as 
rather extreme. Indeed, when $b_m<0$, the dispersion 
relation becomes complex which causes no problem since this situation 
can even occur in classical physics. However, the initial state is 
chosen in this region. Since it is the state which minimizes the 
energy and since the energy might be unbounded from below in this 
region, this choice could be problematic. 
\par
This leads us to the idea that we should explore a case 
in which the dispersion relation lies between the standard relation 
and Unruh's one in Fig. (\ref{disp}). In this case, the 
alteration of the dispersion relation does not have the two 
shortcomings mentioned before. Typically, such a relation is 
given by
\begin{equation}
\label{displog}
\omega =\frac{k_{\rm C}}{\gamma }\ln \biggl(1+\gamma \frac{k}{k_{\rm C}}\biggr),
\end{equation}
where $\gamma $ is a free parameter. Interestingly enough, the case $\gamma =1$ 
has already been studied by Kowalski-Glikman \cite{KG} 
for completely different reasons. The adiabaticity coefficient is given by
\begin{equation}
\label{allog}
\alpha _{\rm KG}(n,\eta )=\frac{\lambda }{\gamma l_{\rm C}}
\biggl(1+\gamma \frac{l_{\rm C}}{\lambda }\biggr)\ln \biggl(
1+\gamma \frac{l_{\rm C}}{\lambda }\biggr).
\end{equation}
When $l_{\rm C}\gg \lambda $, $\alpha _{\rm KG}$ tends very slowly to 
infinity with a rate which depends on $\gamma $. Therefore, 
it seems that adiabaticity is preserved and that we should not expect 
any change in the final spectrum. This is also the result reached by 
the study done in Ref. \cite{KG}.
\par
The conclusion of this work is mitigated. The predictions of inflation 
appear less robust than those of black hole physics. It is 
possible to find cases where adiabaticity is violated and where the 
modifications to the usual power spectrum can be explicitly calculated, see 
Refs. \cite{MB,BM}. However, these modifications can be considered as very 
strong and each time one tries to introduce less drastic modifications, the 
final result is the same as for the unmodified dispersion relation. 
In a certain sense, the presently available data 
already constrains the physics beyond the Planck scale. Therefore, it is
reasonable to conjecture that the allowed region of monotonic dispersion 
relations (the present study does not constrain oscillatory dispersion 
relations) lies roughly between the standard dispersion relation and the Unruh 
dispersion relation, i.e. corresponds to the dashed region in 
Fig. (\ref{disp}). In fact, when the dispersion relation lies
slightly lower than Unruh's relation, one should
still have no modifications of the spectrum and therefore, 
the dashed region can be considered as a lower bound on the region for 
unmodified spectra.

\end{document}